\documentclass{article}      
\usepackage{graphicx}
\usepackage{epstopdf}
\title{Electromagnetic Form Factors and the Hypercentral CQM}  
\author{M. De Sanctis \\
INFN,Sezione di Roma1, \\
Piazzale Aldo Moro, Roma (Italy)\\
and Universidad Nacional de Colombia, Bogot\`{a} (Colombia)\\
M.M. Giannini, E. Santopinto, A. Vassallo\\
Universit\`a di Genova and INFN, Sezione di Genova,\\
via Dodecaneso 33, 16146 Genova (Italy)\\
}

\date{}
\begin{document}

\maketitle  

\begin{abstract}
We present new results concerning the electromagnetic form factors of the
nucleon using a semirelativistic version of the hypercentral
Constituent Quark Model and a relativistic current. The calculations,
performed without free parameters, provide an overall description of the
form factors, with some difficulty for the neutron charge distribution.
The complex structure of the constituent quarks is taken into account implicitly
introducing phenomenological constituent quark form factors. In this way, 
 a detailed reproduction of the experimental data up to $5~GeV^2$ is obtained.
\end{abstract}

The new data on the ratio of the electric and magnetic form factors of the
proton \cite{Jones:1999,Gayou:2001,Gayou:2002}, showing an unexpected
decrease
with $Q^2$, have again triggered interest in the description
of the internal nucleon structure in terms of various effective
models: bag models, chiral soliton models, constituent
quark models, etc.$~$.

In 1973 Iachello {\em et al.} \cite{Iachello:1972nu} were able to
obtain a good reproduction of all the existing nucleon form factor data
using
a Vector Meson Dominance
(VMD) model introducing an
intrinsic form factor to describe the internal structure of the nucleon.
If the ratio $G_E/G_M$ is plotted, the results of the original fit
decrease
with $Q^2$ and cross zero at about $8~GeV^2$.
In 1996 Holzwarth \cite{Holzwarth} showed
that the simple Skyrme soliton model, with vector meson corrections and
with
the initial and final nucleon states boosted to the
Breit Frame, leads to a $G^p_E$ that decreases with $Q^2$ and
crosses zero at $10~GeV^2$.
In this case the crossing is due to a zero in the Skyrme model form factor
as
it is also explained in Ref.~\cite{Holzwarth02}.
In 2000 Cardarelli and Simula \cite{Cardarelli:2000} using light cone
constituent quark models extracted $G^p_M$ from the matrix elements of the
y-component of the current, and showed that the decreasing of the ratio is
due
to the Melosh rotations.  In 1996, Frank {\em et al.}
\cite{Frank:1995pv} have constructed a relativistic light cone
constituent quark model and calculated the electric and magnetic form
factors
of the proton. If their calculations are plotted as a ratio of
the electric and magnetic form factors, one can see a strong decrease
with $Q^2$ due to the presence of a zero in the
electric form factor at $Q^2~=~6~GeV^2$ \cite{Miller:2002}.
In 1999~ with the hypercentral Constituent Quark Model (hCQM)~
\cite{pl,sig}, boosting the initial and final state to the Breit Frame and
considering relativistic corrections to the non relativistic current
\cite{mds}, we showed explicitly that the decrease is a relativistic
effect and it disappears without these corrections
\cite{ts99,rap,fra99}. This calculation made use of the nucleon form
factors previously determined in \cite{mds}. Using a chiral CQM and a
point
form dynamics, the Pavia-Graz group \cite{Wagenbrunn:2000es,Boffi:2001zb}
obtained
a good reproduction of the form factors up to $4~GeV^2$ and a decrease of
the ratio.
A similar reproduction \cite{Merten:2002nz}
has also been obtained within a Bethe Salpeter approach to a constituent
quark model
with an instanton based interaction \cite{Loring:2001kx}. Using the MIT
Bag
model, a sharp decrease with $Q^2$ of the ratio is expected and a change
of
sign at $Q^2~=1.5~GeV^2$. The inclusion of the pion cloud not only
improves the
static properties of the model and restores the chiral symmetry, but also
the behaviour of the ratio $G^p_E/G^p_M$
\cite{Lu:1997sd,Lu:1999np,Lu:yc}. Lattice QCD calculations extrapolated to
the chiral
limit \cite{Thomas4} give rise to interesting results. Finally we
can say that the new VMD fit by Iachello and Wan \cite{Iachello2004} and
Iachello and Bijker \cite{Bijker:2004yu}, the
extended VMD model by Lomon \cite{Lomon:2002jx}, the soliton
model calculation by Holzwarth \cite{Holzwarth,Holzwarth02},
the calculation by Miller \cite{Miller:2002}
 and the relativistic two quark spectator model calculation by Ma \emph{et
al.}\cite{Ma} describe the new Jlab data quite well. For reviews on the
subject the readers are referred to ~\cite{Gao:2003ag,dejager}.

Here, we present new results
obtained with the hCQM, using a semirelativistic Hamiltonian and a
relativistic quark current. Preliminary
results have been presented at various Conferences
\cite{trieste,tasco,baryon}.

First, we briefly review the non relativistic hCQM \cite{pl}.
The experimental $4-$ and $3-$ star non strange resonances can be arranged
in  $SU_{sf}(6)$ multiplets. This means that the quark dynamics
has a dominant spin-flavour invariant part, which accounts for the average
multiplet energies. In the hCQM it is assumed to be
\cite{pl}
\begin{equation}\label{eq:pot}
V(x)= -\frac{\tau}{x}~+~\alpha x,
\end{equation}
where $x$ is the hyperradius
\begin{equation}
x=\sqrt{\mbox{\boldmath $\rho$}^2+{\mbox{\boldmath $\lambda$}^2}}~~,
\end{equation}
with {\boldmath $\rho$} and {\boldmath $\lambda$} being the Jacobi
coordinates
describing the internal quark motion. \\
\noindent Interactions of the linear plus Coulomb-like type have been used
for long time for the meson sector, e.g. the Cornell potential. Moreover
this
form has been supported by recent Lattice QCD calculations \cite{bali}.\\
\noindent In the case of
baryons, a so called hypercentral approximation has been
introduced \cite{has,rich}; this approximation amounts to average
any two-body
potential for the three quark system over the hyperangle
$t=arctg(\frac{{\rho}}{{\lambda}})$ and the angles $\Omega_{\rho}$ and
$\Omega_{\lambda}$, and it works quite well, especially for the lower part
of the
spectrum \cite{hca}.  In this respect, the hypercentral potential of
Eq.(\ref{eq:pot}) can be considered as the hypercentral approximation of
a two-body linear plus Coulomb-like potential.
The splittings within the multiplets are produced by a perturbative term
breaking the $SU_{sf}(6)$ symmetry,
which, as a first approximation, can be assumed to be the
standard hyperfine interaction $H_{hyp}$ \cite{is}.

In the baryon rest frame, the three quark hamiltonian, in the non
relativistic case,  can be written as \cite{pl}:
\begin{equation}\label{eq:ham}
H_{NR} = 3m+\frac{\mbox{\boldmath $p$}_\rho^2+
\mbox{\boldmath $p$}_\lambda^2}{2m}-\frac{\tau}{x}~
+~\alpha x+H_{hyp},
\end{equation}
where $m$ is the quark mass (taken equal to $1/3$ of the nucleon mass) and
{\boldmath $p$}$_\rho$, {\boldmath $p$}$_\lambda$ are the conjugate
momenta of the Jacobi coordinates.
We construct the non strange baryons as bound states of three constituent
quarks,
taking properly into account the antisymmetrization with respect to all
quark
coordinates. Because of the  hyperfine mixing, each baryon state is a
superposition
of various $SU(6)$ configurations. The hamiltonian  (\ref{eq:ham}) is
diagonalized
in the space of the baryon rest frame states.
The strength of the hyperfine interaction is determined in order to
reproduce the
$\Delta-N$ mass difference and the remaining two free parameters are
fitted to the spectrum
leading to the values $\alpha= 1.61~fm^{-2}$ and $~\tau=4.59~$ \cite{pl}.
 Keeping these parameters fixed, this non relativistic constituent quark
model
has been used to calculate
various physical quantities of interest: photocouplings
\cite{aie},
electromagnetic transition amplitudes \cite{aie2} and, introducing
relativistic
corrections to the one-body non relativistic current, also the elastic
nucleon
form factors \cite{mds} and the ratio between the electric and magnetic
form
factors of the proton \cite{rap}. We have shown that kinematical
relativistic
corrections (such as boosts and a relativistic one-body current with an
expansion
in the quark momenta up to the first order, keeping the exact dependence
on
the  momentum transfer $Q^2$)  are very important for the elastic form
factors
\cite{rap,mds}
but yield only minor corrections in the transition ones \cite{mds2}.

We propose a semirelativistic hypercentral constituent quark
model based on the following hamiltonian \cite{traini}
\begin{equation}\label{eq:hrel}
H = \sum_{i=1}^3 \sqrt{{\vec{k}_{i}}^2+m^2}-\frac{\tau}{x}~
+~\alpha x+H_{hyp}.
\end{equation}
which employs the relativistic kinetic energy, where $\vec{k}_{i}$ are the
quark
three momenta in the rest frame ($\sum_{i=1}^3 ~\vec{k}_i~= ~ 0$).
The hamiltonian of Eq. (\ref{eq:hrel}) is solved by means of a variational
method
(a complete description of the variational solution of this equation,
using the
hyperspherical formalism, will
be published elsewhere \cite{traini}).
The resulting spectrum is not much different from the non relativistic
one and the parameters $\alpha$ and $\tau$ of the potential are only
slightly modified,
while the constituent quark masses are $m=100~MeV$.

The Hamiltonian \ref{eq:hrel} can be used within a covariant approach if a
Bakamjian-Thomas (BT) construction \cite{BT} is performed. In the BT method the
interaction
is introduced by adding to the free mass operator
$M_0~=~\sum_{i=1}^3~\sqrt{{\vec{k}_i}^2+m^2}$ an interaction term $M_I$, 
in such a way
that the total mass $M ~= ~M_0 + M_I$ commutes with the Poincar\'{e} 
generators \cite{KP} 
A complete set of Poincar\'e generators can be built according to the
prescriptions provided by the point form approach; in this way
the $4$-momentum operators $P_{\mu}$ contain interactions while the
rotations and the Lorentz boosts are interaction free \cite{kl1}. 
The general 3-quark state is defined on the product space of the
one-particle spin-$1/2$, positive energy representation of the
Poincar\'{e} group \cite{Wagenbrunn:2000es}.
The rest frame free states can be written as
\begin{equation} \label{eq:free}
|\mbox{\boldmath $k$}_1,
 \mbox{\boldmath $k$}_2,\mbox{\boldmath $k$}_3,>÷=÷
u(\mbox{\boldmath $k$}_1)
    u(\mbox{\boldmath $k$}_2)u(\mbox{\boldmath $k$}_3)
\end{equation}
where $u(\mbox{\boldmath $k$}_i)$ is the positive energy Dirac spinor of 
the i-th quark
and the three-momenta satisfy the condition $\sum_{i=1}^3 ~\vec{k}_i~= ~ 0$.
In the rest frame of the three quark system, the stationary part of the
equation $P_{\mu} ÷|\Psi >÷=÷ p_{\mu} ÷÷|\Psi >$ is identified with the 
eigenvalue
problem corresponding to the hamiltonian \ref{eq:hrel}. The nucleon 
state in the
space provided by the states of Equation ( \ref{eq:free}) is then given by 
\begin{equation}
  \Psi(\mbox{\boldmath $k$}_1,
 \mbox{\boldmath $k$}_2,\mbox{\boldmath $k$}_3)=
u(\mbox{\boldmath $k$}_1)
    u(\mbox{\boldmath $k$}_2)u(\mbox{\boldmath $k$}_3)
    \varphi(\mbox{\boldmath $p$}_\rho,\mbox{\boldmath $p$}_\lambda)
    \label{eq:relwf}
\end{equation}
where $\mbox{\boldmath $p$}_\rho=(\mbox{\boldmath $k$}_1-
\mbox{\boldmath $k$}_2)/\sqrt{2}$ and $\mbox{\boldmath $p$}_\lambda=
(\mbox{\boldmath $k$}_1+
\mbox{\boldmath $k$}_2-2\mbox{\boldmath $k$}_3)/\sqrt{6}$ are the Jacobi momenta
calulated from the rest frame quark momenta $\vec{k_i}$ and 
$\varphi(\mbox{\boldmath
$p$}_\rho,\mbox{\boldmath $p$}_\lambda)$
 is the eigenfunction of the Hamiltonian of equation (\ref{eq:hrel}).
In order to perform the transformation to a different reference frame, 
we introduce the
velocity states \cite{kl1}
\begin{equation}
|v,\mbox{\boldmath $p$}_1,\mbox{\boldmath$p$}_2,\mbox{\boldmath $p$}_3>÷=÷
U_{B(v)} ÷|\mbox{\boldmath $k$}_1,\mbox{\boldmath $k$}_2,\mbox{\boldmath $k$}_3>
\end{equation}
where $U_{B(v)}$ is a Lorentz boost corresponding to the velocity $v$ and
$p_i^{\mu}$ are the quark momenta in the trasformed frame.
We apply to each quark spinor a canonical boost, obtaining that 
the transformed
quark momenta $p_i^{\mu}$ satisfy the relation 
\begin{equation}
\sum_{i=1}^3 ~p_i^{\mu}~=
~\frac{P^{\mu}}{M}÷ \sum_{i=1}^3 ~\epsilon (\vec{k_i}), 
\end{equation}
where $\epsilon (\vec{k_i})$ is 
the rest frame quark energy, $ P^{\mu}$ is the observed nucleon 4-momentum and $M$
its mass. 
Moreover, 
$p_i^{\mu}÷= B(v) k_i$. Having applied  canonical boosts, the conditions for a Point form 
approach \cite{kl1,meld} are satisfied. In particular, the three quark perform the same rotation 
and the quark spins can be coupled as in the nonrelativistic case \cite{kl1,kl2}.

We now proceed to calculate the elastic nucleon form factors. 
We choose to work in the Breit frame, where the initial and final states acquire a
momentum along the $z$-axis $p_z$ and $pÕ_z$, respectively, with $p_z=-pÕ_z=-q/2$,
$q$ being the $z$-component of the virtual photon momentum. 

The nucleon electromagnetic form factors can be extracted from the 
matrix elements of the nucleon electromagnetic current between the initial and
final nucleon states of eq. \ref{eq:relwf}, according to the formalism described 
in Ref. \cite{kl1}.

The  current operator is written in impulse approximation, \emph{i.e.} 
it is chosen to be the sum of the single quark currents 
\cite{mds,rap,Wagenbrunn:2000es,Boffi:2001zb}; the matrix elements of the quark 
current in the space of the single quark free spinor states are given by

\begin{eqnarray}\label{eq:curr}
\langle p_1, p_2, p_3 \vert J_{\mu}  \vert p_1', p_2', p_3'
\rangle~=~~~~~~~~~~~~~~~~~~\nonumber \\
  \sum_i \bar{u}_i (p_i) J_{i\mu} u_{i} (p_i') \\
\bar{u}_j (p_j) u_{j} (p_j')\delta (p_j -p_j') 
\bar{u}_k (p_k) u_{k} (p_k')\delta (p_k -p_k')  \nonumber
\end{eqnarray}

where $i,j,k$ is an even permutation of the indexes $1,2,3$ and
\begin{equation}\label{eq:bare}
\bar{u}_i (p_i) J_{i\mu} u_{i} (p_i')~=
~\bar{u}_i (p_i) e_i 
\gamma_\mu(i)   u_{i} (p_i')
\end{equation}
where $e_i$ is the quark charge and $Q^2$ is the virtual photon 
squared tetramomentum.

The single quark current is covariant but in principle not conserved,
however a 
conserved current can be obtained by the simple transformation 
$j'_{\mu} =j_{\mu} - q_{\mu}
(q \dot j)/q^2$, 
where $q_{\mu}$ is the virtual photon tetramomentum; 
choosing the z-axis along the space component of $q_{\mu}$, such procedure does not affect the 
$0,1,2$ components of the current, from which the elastic form factors are extracted.

The nucleon matrix elements are then calculated making use of the wave
functions of eq. (\ref{eq:brwf}) and are given by

\begin{eqnarray}
\label{eq:nuc}
J^N_{\mu} ~=~ \frac{3}{J} \int \frac{d^3p_1}{\epsilon(\vec{p}_1)}
\frac{d^3p_2}{\epsilon(\vec{p}_2)} ~\bar{\Psi}(\vec{p}_1,
\vec{p}_2,\vec{P}_F) ~~~~~~~~~\nonumber\\~
f(\vec{p}_1,\vec{p}_2,\vec{P}_F) ~e_3
~(F_1^q(Q^2)+F_2^q(Q^2))\gamma_\mu(3)   \nonumber\\
-\frac{1}{2m} (p_3+p_3')_\mu F_2^q(Q^2) ~f(\vec{p}_1,\vec{p}_2,\vec{P}_I) 
~ \Psi(\vec{p}_1, \vec{p}_2,\vec{P}_I) 
\end{eqnarray}
where the factor $3$ accounts for the symmetry of the wave function, $J$
is the Jacobian for the transformation from the single quark to the
Jacobi coordinates, $f(\vec{p}_1,\vec{p}_2,\vec{P})$ is a normalization
factor which ensures the correct charge normalization of the current.

\begin{figure*}[!ht]
  \centering
  \includegraphics[width=8.cm]{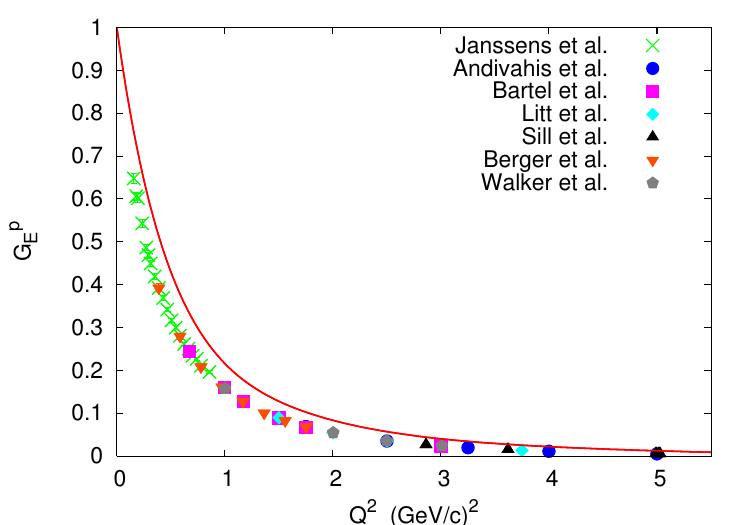}
  \includegraphics[width=8.cm]{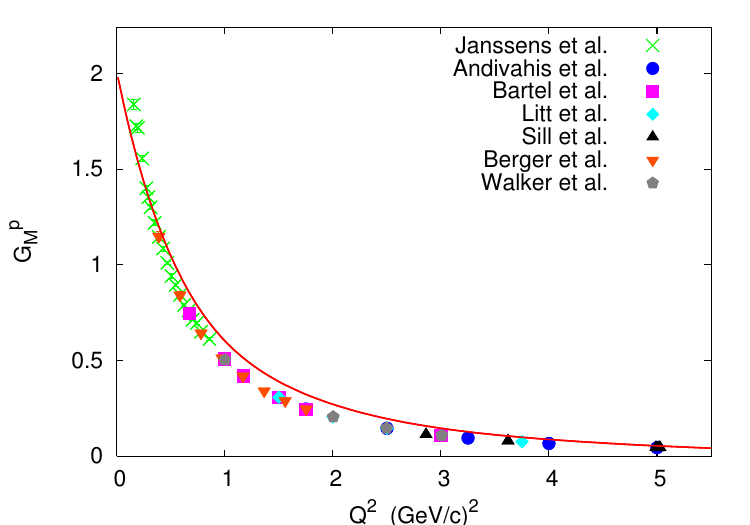}\\
  \includegraphics[width=8.cm]{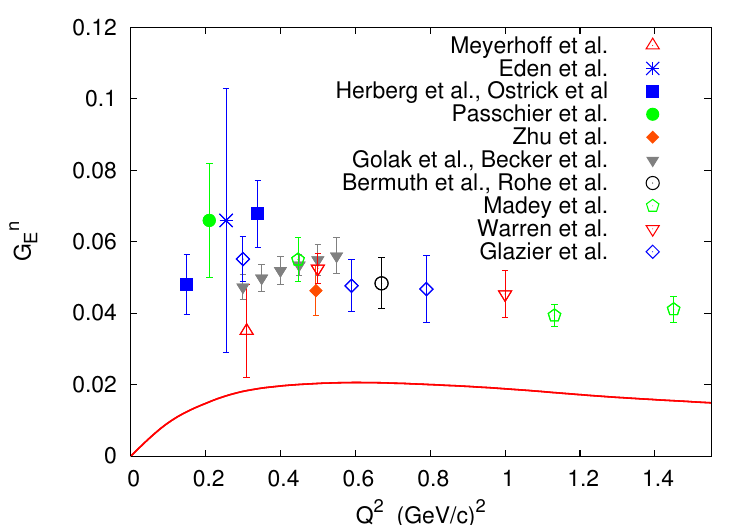}
  \includegraphics[width=8.cm]{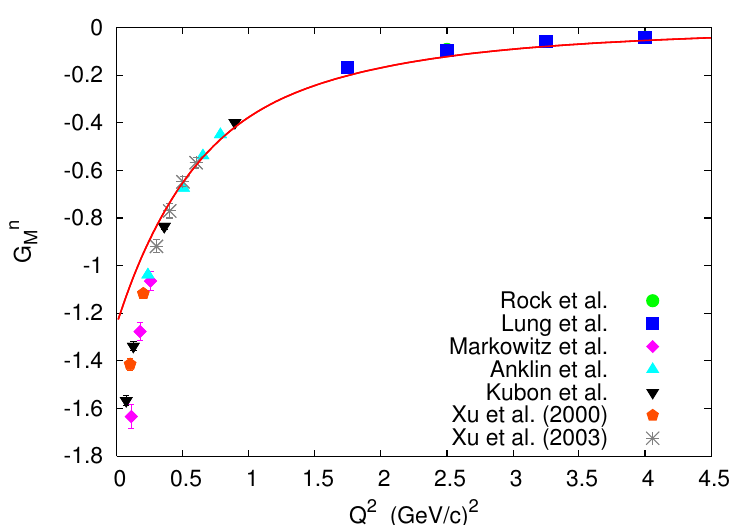}
\caption{ (Color online) Elastic form factors of the nucleon. The solid
line
    corresponds to the semirelativistic hCQM calculation with the quark
current of Eq. (\ref{eq:bare}). The
experimental
    data for $G_M^p$ are taken from the reanalysis made by Brash {\em et
al.} \cite{Brash:2004} of the data from \cite{Andivahis:1994,Bartel:1973,
      Janssens:1966,Litt:1970,Sill:1993,Berger:1971,Walker:1994}; the
points shown for $G_E^p$ are obtained from the data on $G_M^p$ and the
linear fit \cite{Brash:2004} of the Jlab data on the $\mu_pG_E^p/G_M^p$
ratio; for $G_E^n$ the experimental data are taken from
\cite{Meyerhoff:1994,Eden:1994,Herberg:1999,Passchier:1999,Zhu:2001,
Golak:2001,Bermuth:2003,Madey:2003,Warren:2004,Glazier:2005}, and for 
$G_M^n$ the experimental data are taken from
    \cite{Anklin:1998,Kubon:2002,Lung:1993,Markowitz:1993,Rock:1982,
      Xu:2000,Xu:2003}.}
  \label{fig:ff_nudo_mds}
\end{figure*}

\begin{figure}[!ht]
  \begin{center}
    \includegraphics[width=8cm]{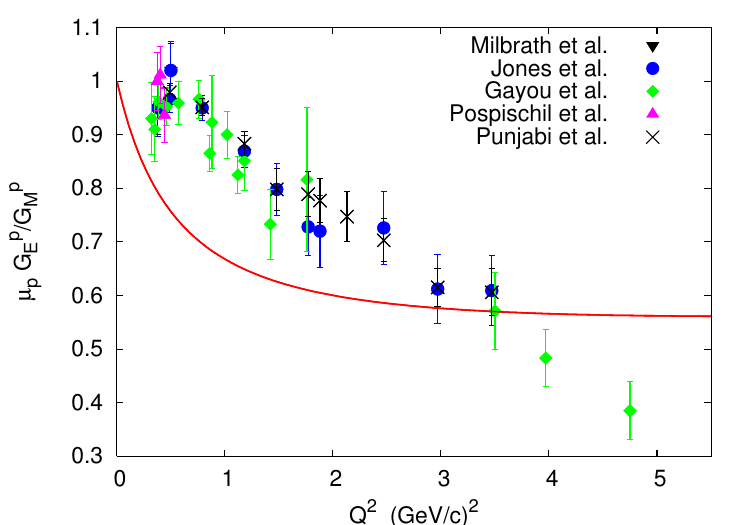}
    \caption{ (Color online) The ratio $\mu_p G_E^p/G_M^p$ from
polarization
      transfer compared with the semirelativistic hCQM calculation
      with the quark
current of Eq. (\ref{eq:bare}) (solid line). The experimental data are
taken from ~\cite{Milbrath99,Jones:1999,Gayou:2001,Gayou:2002,Pospischil03,
Punjabi:2005}.}
    \label{fig:rap_nudo_mds}
  \end{center}
\end{figure}

The resulting theoretical form factors of the nucleon
can be seen in Figs. \ref{fig:ff_nudo_mds} and \ref{fig:rap_nudo_mds}. The
results
reported in Figure \ref{fig:ff_nudo_mds} show a quite good reproduction of
the data
even if some problems are still present especially at low $Q^2$.
Nonetheless
there is a great improvement in comparison with the non relativistic
calculations of Refs \cite{mds,rap}, where an expansion in the quark
momentum was performed.
Moreover another important improvement is given by the use of
semirelativistic wave functions obtained from the hamiltonian
(\ref{eq:hrel}).

Constituent quarks can be in principle considered as
composite objects \cite{petronzio} and accordingly we parametrize
phenomenologically their
structure by means of constituent quark form factors, as already done by
other
authors \cite{Cardarelli:1995dc,petronzio}. The matrix elements of the
quark current \ref{eq:bare} are substituted with the following ones

\begin{eqnarray}\label{eq:qff}
\bar{u}_i (p_i) J_{i\mu} u_{i} (p_i')~=~ ~~~~~~~~~~~~~~\nonumber \\
\bar{u}_i (p_i) e_i
~(F_1^q(Q^2)+F_2^q(Q^2))
\gamma_\mu(i) \\
- \frac{1}{2m}
    (p_i+p_i')_\mu F_2^q(Q^2)  u_{i} (p_i') \nonumber
\end{eqnarray}
where $F_1^q(Q^2)$ and $F_2^q(Q^2)$ are, respectively, the Dirac and Pauli
quark form factors. 
We choose the $Q^2$ behavior of the constituent quark
form factors as a linear combination of monopole and dipole.

\begin{figure*}[!ht]
  \centering
  \includegraphics[width=8cm]{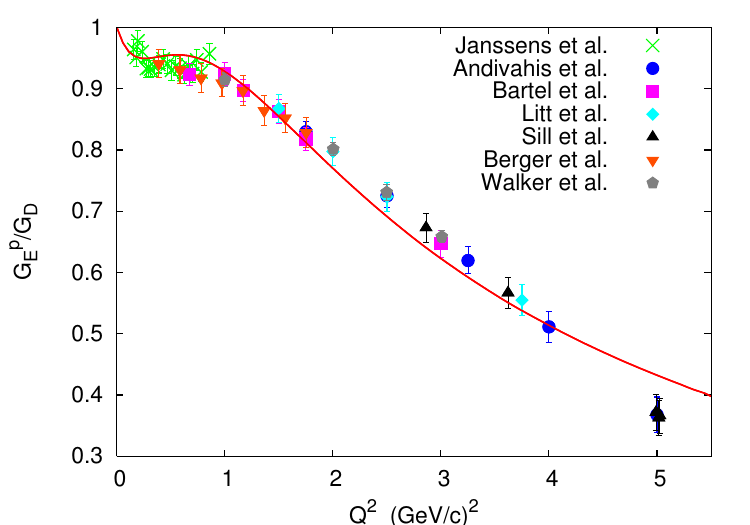}
  \includegraphics[width=8cm]{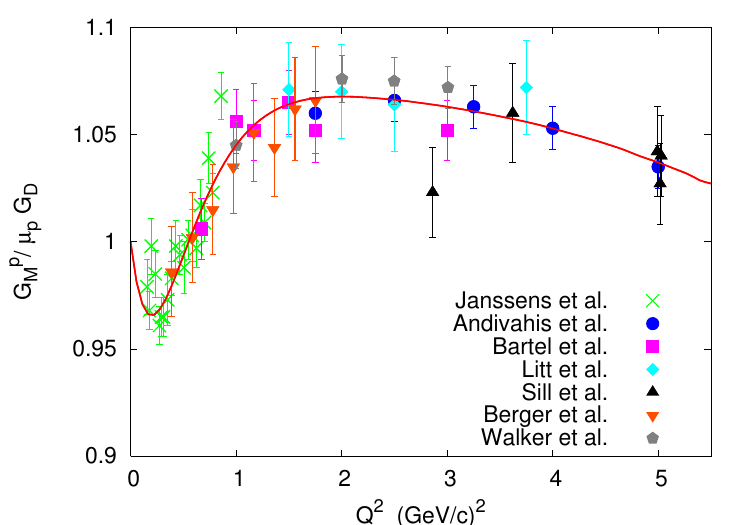}\\
  \includegraphics[width=8cm]{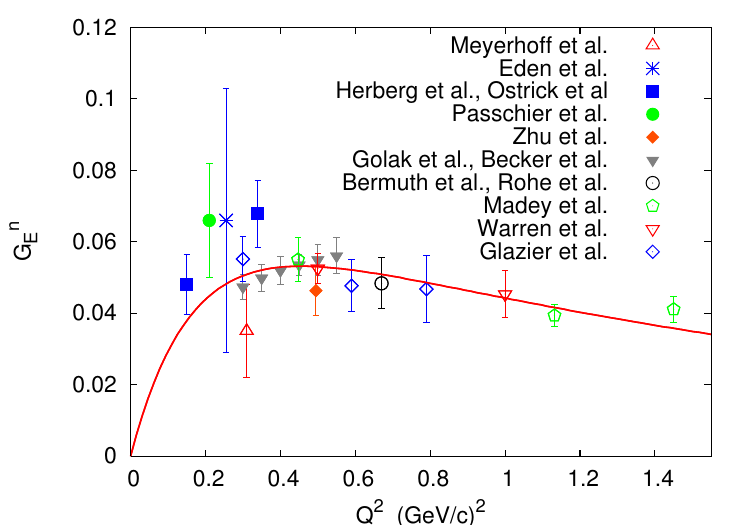}
  \includegraphics[width=8cm]{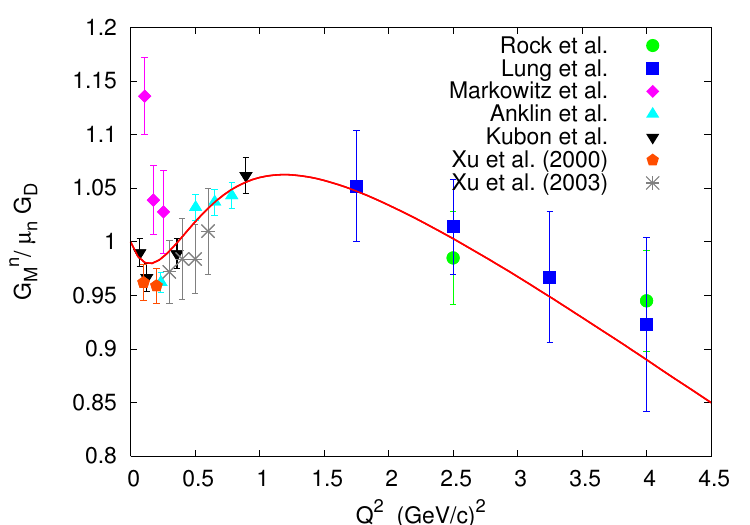}
  \caption{(Color online) Elastic form factors of the nucleon. The solid
line
    corresponds to the semirelativistic hCQM calculation with constituent
    quark form factors. The experimental
    data for $G_M^p$ are taken from the reanalysis made by Brash {\em et
al.} \cite{Brash:2004} of the data from \cite{Andivahis:1994,Bartel:1973,
      Janssens:1966,Litt:1970,Sill:1993,Berger:1971,Walker:1994}; the
points shown for $G_E^p$ are obtained from the data on $G_M^p$ and the
linear fit
    \cite{Brash:2004} of the Jlab data on the $\mu_pG_E^p/G_M^p$ ratio;
    for $G_E^n$ the experimental data
    are taken from \cite{Meyerhoff:1994,Eden:1994,Herberg:1999,
      Passchier:1999,Zhu:2001,Golak:2001,Bermuth:2003,Madey:2003,
      Warren:2004,Glazier:2005}, and for $G_M^n$ the experimental data are
    taken from
    \cite{Anklin:1998,Kubon:2002,Lung:1993,Markowitz:1993,Rock:1982,
      Xu:2000,Xu:2003}.}
  \label{fig:ff_comp}
\end{figure*}

\begin{figure}[!ht]
  \begin{center}
    \includegraphics[width=8cm]{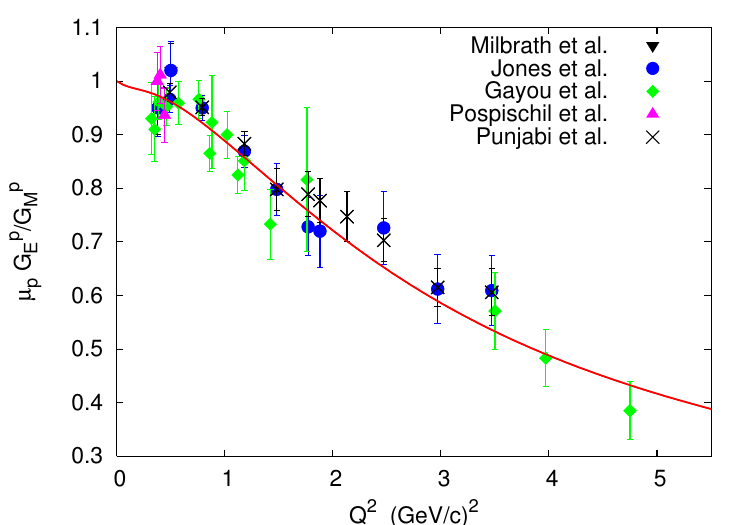}
    \caption{(Color online) The ratio $\mu_p G_E^p/G_M^p$ from
polarization
      transfer compared with the semirelativistic hCQM calculation with
      constituent quark form factors (solid line).
      The experimental data are taken from~\cite{Milbrath99,Jones:1999,
        Gayou:2001,Gayou:2002,Pospischil03,Punjabi:2005}.}
    \label{fig:rap_comp}
  \end{center}
\end{figure}

By fitting the free parameters to the reproduction of $G_M^p$, $G_M^n$,
$G_E^n$
and $\mu_pG_E^p/G_M^p$ we obtain the curves shown in Fig.
\ref{fig:ff_comp} and
\ref{fig:rap_comp}. The very recent data on the ratio from Jlab
\cite{Punjabi:2005}
are also reported in Fig. \ref{fig:rap_comp} for completeness, even if
they have not
yet been included in the fitting procedure.

As it can be seen in Fig. \ref{fig:ff_comp} and  \ref{fig:rap_comp}
the experimental data are very well reproduced. The goodness of the
reproduction is emphasized by plotting the form factors divided by the
dipole
form.  With respect to the non relativistic case, the semirelativistic
wave
functions have more high momentum components.
This fact, together with the application of exact boosts to the Breit
Frame,
leads to an improvement in the reproduction of the existing data on the
electromagnetic form factors.
However,  a good description of the data is obtained only if
phenomenological
constituent quark form factors are introduced in the electromagnetic
current.
In this way we have a  very nice agreement with the available experimental
data
up to $5~ GeV^2$ .

Finally, it results that for a good reproduction of the elastic form
factor data both the relativistic effects and the composite nature of the
constituent quarks have to be taken into account.
We observe that such constituent quark form factors actually parametrize
not
only  the constituent quark structure but also the relativistic effects
which have not yet been explicitly included in our calculations.

\end{document}